\documentclass[twocolumn,prl,showpacs]{revtex4}%
\usepackage{amsmath}
\usepackage{amsfonts}
\usepackage{amssymb}
\usepackage{graphicx}%
\providecommand{\U}[1]{\protect\rule{.1in}{.1in}}
\begin{document}
\title{Decorrelation of samples in Quantum Monte Carlo calculations and scaling of
autocorrelation time in Li and H$_{2}$O clusters}
\author{D. Nissenbaum, B. Barbiellini and A. Bansil}
\affiliation{Physics Department, Northeastern University, Boston MA 02115}
\date{\today}

\pacs{02.70.Ss, 71.10 -w, 71.15.Nc}

\begin{abstract}
We have investigated decorrelation of samples in Quantum Monte Carlo (QMC)
ground-state energy calculations for large Li and H$_{2}$O nanoclusters.
Binning data as a way of eliminating statistical correlations, as is the
common practice, is found to become increasingly impractical as the system
size grows. We demonstrate nevertheless that it is possible to perform
accurate energy calculations$-$without decorrelating samples$-$by exploiting
the scaling of the integrated autocorrelation time $\tau$ as a function of the
number of electrons in the system.

%The scaling factor for $\tau$ adds an equivalent factor to the scaling of the total computation time - a factor which has not been considered previously.

%In large nanoclusters binning the data is found not to eliminate %statistical correlations in a reasonable amount of computation time.

%I have shortened the abstract somewhat to make for less repetition and easier reading--ARUN

\end{abstract}
\maketitle

Quantum Monte Carlo (QMC) methods are becoming increasingly important in
electronic structure calculations for a variety of systems of current interest
\cite{review, review2}.\ Although effective QMC methods have been developed in
recent years to optimize the parameters of many-body wavefunctions
\cite{umrigar88,sga,rappe,bressanini,sorella,filippi}, QMC calculations are
fundamentally limited by the statistical accuracy that can be obtained in a
reasonable amount of computation time. Despite the growing maturity of core
QMC techniques in terms of efficiency \cite{umrigar93b,umrigar93},
computations of large clusters are hampered by increasing correlations in
samples with increasing system size. In this connection, scaling of the
computation time required to obtain a given statistical accuracy has been
reported as a function of the atomic number $Z$ \cite{ceperley, hammond, ma},
and also as a function of the number of atoms $M$ in non-metallic clusters
\cite{galli}. Here we discuss the effects of correlated sampling in Li and
H$_{2}$O clusters.

When the number of atoms in a metallic cluster increases, the gap to the
excited states shrinks as the electronic states begin to form a band and
approach the bulk metallic character. This shrinking gap between the ground
state and the excited states induces progressively greater statistical
correlations among samples due to the Markovian character of the QMC sampling
algorithm \cite{footnoteCorrelations}. We address this problem through an
analysis of the integrated autocorrelation time $\tau$ \cite{liu}\thinspace
which has not been considered in previous work on scaling; in particular,
Williamson, \textit{et al}. \cite{galli} do not consider this correlation
factor in their studies of non-metallic clusters. We demonstrate with the
example of Li clusters that the problem of correlations among samples
increases rapidly as the system size grows, making it increasingly difficult
to decorrelate samples and determine the statistical error of the QMC
calculation. Nevertheless, by invoking the scaling relationship of $\tau$,
efficient computations of large clusters become feasible, even in the presence
of highly correlated samples.

The relevant technical details of our calculations are as follows. For a given
many-body wave function, $\Phi_{0}(x)$, we employ Umrigar's modified version
of discrete Langevin dynamics \cite{umrigar93} to generate a sequence of QMC
samples \cite{footnote3}. Our QMC runs were carried out using a modification
of the QMcBeaver code \cite{qmcbeaver}. We chose to move all electrons during
each QMC step rather than performing single-electron moves
\cite{footnoteSingleGlobalExplanation}. In all cases, an acceptance ratio of
50\% was maintained. We note that achieving equilibration for large clusters
is quite demanding. In particular, the Li$_{64}$ and $\left(  \text{H}%
_{2}\text{O}\right)  _{20}$ clusters each required about one week to
equilibrate \cite{footnoteEquilibration}.

After equilibration, the QMC walker's steps follow a probability distribution
$\pi(x)=|\Phi_{0}(x)|^{2}$. The estimate of the total energy $E$=$<\Phi
_{0}|H|\Phi_{0}>$ is obtained from the average of the local energies
$h(x)=H\Phi_{0}(x)/\Phi_{0}(x)$ over the steps taken by the walker. Estimates
of $E$ from multiple independent calculations are distributed according to a
normal distribution, with variance $\sigma^{2}$ estimated by \cite{liu}
\begin{equation}
\sigma^{2}\approx\sigma_{0}^{2}(1+2\sum_{i=1}^{\infty}C_{h}(i)),
\label{var_intermsof_autocorr}%
\end{equation}
where $\sigma_{0}^{2}$ is the estimate of the variance of $E$ calculated as
though the samples were uncorrelated, i.e. $\sigma_{0}^{2}=\sigma_{raw}^{2}%
/N$, where $\sigma_{raw}^{2}$ is the variance of $h(x)$ over $N$ steps.
$C_{h}(i)$, the autocorrelation of $h$ at $i$ steps, is given by $C_{h}\left(
i\right)  =\left\langle h\left(  x^{j}\right)  h\left(  x^{j+i}\right)
\right\rangle -\left\langle h\left(  x^{j}\right)  \right\rangle ^{2}$
(normalized to the point $i=0$), where $j$ is an arbitrary step number in the
post-equilibration phase, and the brackets $\left\langle \ldots\right\rangle $
denote the expectation value over an infinite ensemble. The "integrated
autocorrelation time" $\tau$ is then defined such that the variance
\begin{equation}
\sigma^{2}=\sigma_{raw}^{2}/(N/\tau)\text{,} \label{sigma_intermsof_autocorr}%
\end{equation}
so that the "effective" number of steps taken by the walker is given not by
$N$, but by $N/\tau$. Eq. \ref{sigma_intermsof_autocorr}\ generalizes the
statistical error whose scaling with system size has been discussed by
Williamson, \emph{et al. }\cite{galli} by the addition of the quantity $\tau$,
the factor which describes the effect of statistical correlations between QMC
moves. In the discrete case at hand, we have
\begin{equation}
\tau=(1+2\sum_{i=1}^{\infty}C_{h}(i))~. \label{corrtime_intermsof_autocorr}%
\end{equation}
By comparing Eqs. (\ref{var_intermsof_autocorr}) and
(\ref{corrtime_intermsof_autocorr}), we see that
\begin{equation}
\tau=(\sigma/\sigma_{0})^{2}. \label{tau_eq_ss02}%
\end{equation}
\ 

The ratio in Eq. \ref{tau_eq_ss02} is sometimes also called the "statistical
inefficiency"; see Refs. \cite{FC,KB,AT}. In order to determine the ratio
$(\sigma/\sigma_{0})^{2}$ and thus obtain $\tau$, we employed the recently
developed Dynamic Distributable Decorrelation Algorithm (DDDA) \cite{kent} to
block data efficiently. Like other data blocking algorithms \cite{footnote5},
the DDDA is a renormalization group method that relies on the appearance of a
plateau to determine the correct variance of the mean value of the energy. The
DDDA performs the blocking with nearly zero overhead both in time and in
memory, and for these reasons it is well-suited for our study of the
convergence properties of large systems.

Simulations were carried out on Li clusters with 4, 8, 20, 26 and 64 atoms
\cite{lithium}. Li is an interesting element because it displays subtle
bonding properties \cite{MarxLi,marx}. Moreover, it contains only a few
electrons per atom, which makes it possible to consider relatively large Li
clusters. We have used Hartree-Fock \cite{footnoteHF}\ many-body wave
functions $\Phi_{0}(x)$ constructed with one-particle orbitals obtained with
the software Jaguar \cite{jaguar}.%

%TCIMACRO{\FRAME{ftbpFU}{3.2681in}{2.565in}{0pt}{\Qcb{(Color online) Normalized
%trial standard deviation $\sigma_{t}/\sigma_{0}$ and its error bar for 4 to 64
%atom Li clusters. Samples are decorrelated when $\sigma_{t}/\sigma_{0}$
%displays a plateau, where the height of the plateau is related from Eq.
%\ref{tau_eq_ss02} to the square root of the integrated correlation time, i.e.
%$\sqrt{\tau}=\sigma/\sigma_{0}$. Inset shows the linear scaling of $\sqrt
%{\tau}$ with the number of electrons $N_{el}$ in the system. Error bars on
%data points are comparable to dot sizes.}}{\Qlb{fig__1}}{li_clu.eps}%
%{\special{ language "Scientific Word";  type "GRAPHIC";
%maintain-aspect-ratio TRUE;  display "USEDEF";  valid_file "F";
%width 3.2681in;  height 2.565in;  depth 0pt;  original-width 3.2232in;
%original-height 2.5235in;  cropleft "0";  croptop "1";  cropright "1";
%cropbottom "0";
%filename 'C:/Documents and Settings/Physics_on_Fujitsu/My Documents/_Workspace/MathText/My Published Articles/Decorrelation/Li_clu.eps';file-properties "XNPEU";}%
%}}%
%BeginExpansion
\begin{figure}
[ptb]
\begin{center}
\includegraphics[
height=2.565in,
width=3.2681in
]%
{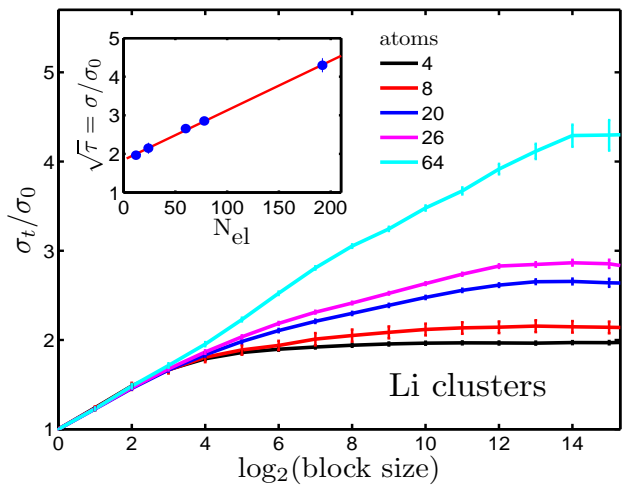}%
\caption{(Color online) Normalized trial standard deviation $\sigma_{t}%
/\sigma_{0}$ and its error bar for 4 to 64 atom Li clusters. Samples are
decorrelated when $\sigma_{t}/\sigma_{0}$ displays a plateau, where the height
of the plateau is related from Eq. \ref{tau_eq_ss02} to the square root of the
integrated correlation time, i.e. $\sqrt{\tau}=\sigma/\sigma_{0}$. Inset shows
the linear scaling of $\sqrt{\tau}$ with the number of electrons $N_{el}$ in
the system. Error bars on data points are comparable to dot sizes.}%
\label{fig__1}%
\end{center}
\end{figure}
%EndExpansion

Fig. \ref{fig__1} shows the value $\sigma_{t}/\sigma_{0}$ plotted as a
function of block size for Li clusters varying in size from 4-64 atoms, where
$\sigma_{t}$ is the trial standard deviation at various block sizes in the
computation. The horizontal axis gives the number of adjacent QMC samples
(block size) pulled piecewise from the full set of samples and aggregated,
creating effective samples which are used in the calculation of $\sigma_{t}$.
The appearance of a plateau, or flat region, in the $\sigma_{t}/\sigma_{0}$
plot indicates full decorrelation of samples. The height of this plateau
yields the true value of the normalized standard deviation $\sigma/\sigma_{0}%
$, which from Eq. \ref{tau_eq_ss02} gives the correct value of the square root
of the integrated correlation time, $\sqrt{\tau}$. For example, for the 4 atom
cluster with $N_{e}$=12 electrons, the $\sigma_{t}/\sigma_{0}$ curve (black
line) flattens to a value of about 2, which is the value shown for $\sqrt
{\tau}$ in the inset. Similarly, the plateau for 20 Li atoms with 60 electrons
(deep blue line) appears for $\sigma_{t}/\sigma_{0}$=2.7.

An increase in the height of the plateau, i.e., in the value of $\sigma
/\sigma_{0}$ or equivalently of $\sqrt{\tau}$, with cluster size is evident in
Fig. \ref{fig__1}. More quantitatively, the inset reveals a linear
relationship of $\sigma/\sigma_{0}$ with the number of electrons $N_{el}$ in
the cluster. It thus follows that the autocorrelation time diverges
quadratically with system size. We emphasize that the total number of points
$n_{p}$ needed to reach the plateau for a given cluster size is much larger
than the corresponding value of $\tau$. For example, for the 64 atom cluster
the plateau appears in Fig. \ref{fig__1} at the value 14 or equivalently for
$n_{p}=2^{14}$, while $\tau=(4.3)^{2}=18.5$. In other words, one would need to
skip $2^{14}$ steps in this case if one wanted to decorrelate samples. Note
that $\tau$ is an integrated correlation time; therefore, skipping $\tau$
steps will \textit{not} produce decorrelated samples. Clearly, because $n_{p}$
grows so large, calculation of the error using reblocking rapidly becomes
intractable for large clusters.

The total time of a computation can be written as $T=t_{s}N$, where $t_{s}$ is
the time per step and $N$ is the total number of steps. From Eq.
\ref{sigma_intermsof_autocorr}, we see that in order to maintain a given
accuracy $\sigma$ for QMC calculations of different clusters, the number of
steps $N$ must change to balance changes in $\sigma_{raw}^{2}$ and $\tau$. For
quadratically scaling $\tau$, the number of steps $N$, and therefore the total
computation time, will have an additional scaling factor of up to $M^{2}$,
where $M$ denotes the number of atoms in the cluster, in order to compensate
for the scaling of $\tau$. This additional scaling factor for $\tau$ is
present whether the quantity being calculated is the energy for the entire
system, or the energy per atom. The scaling factors due to $\sigma_{raw}^{2}$
and $t_{s}$ have been considered in previous work \cite{galli} and are not
considered in this study.

A QMC calculation is complete only when accompanied by a measure $\sigma$ of
its statistical uncertainty, which is given by the onset of the decorrelation
plateau in a plot such as that of Fig. \ref{fig__1}. The failure to obtain
this onset indicates that it is impossible to decorrelate samples during the
run. This "stickiness" of correlated samples in QMC runs impedes our ability
to determine the statistical error $\sigma$ of the computation. Often, in
order to decorrelate samples, the problem of stickiness is overcome by
obtaining a large number of sample points and discarding intermediate
correlated points. However, as we have already pointed out above, it becomes
increasingly difficult in large clusters to obtain the plateau and decorrelate
samples. It may in fact be impossible to decorrelate even a single sample
during the run. In sharp contrast, our procedure requires that all steps are
sampled, with no steps discarded, in order to first calculate $\sigma_{0}^{2}%
$, and then obtain $\sigma^{2}$ by rescaling with an extrapolated value of
$\tau$. In this way, our scheme provides a route for circumventing the problem
of decorrelating samples in large clusters.

As an example of an energy calculation obtained without observing the onset of
a decorrelation plateau, we were able to obtain the energy (per atom) for the
Li$_{64}$ cluster to within $\sigma=0.1$ eV ($\pm10\%$) using only $10^{5}$
QMC steps \cite{footnoteHardware}. We have verified our estimate of the
accuracy of $\sigma$ for this energy calculation by performing multiple,
independently equilibrated runs and observing the spread of the energy results.

\begin{figure}
[ptb]
\begin{center}
\includegraphics[
height=2.8781in,
width=3.493in
]%
{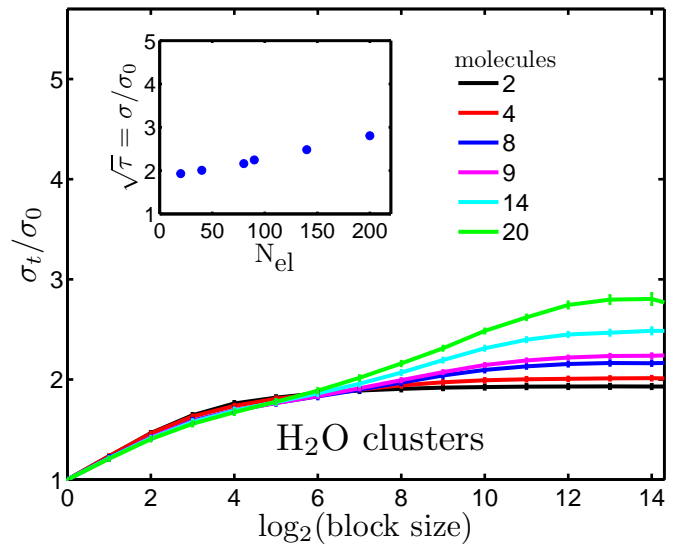}%
\caption{(Color online) Same as Fig. \ref{fig__1}, except that this figure
refers to H$_{2}$O clusters.}%
\label{fig_2}%
\end{center}
\end{figure}

We turn now to consider H$_{2}$O clusters, where we have studied clusters of
2, 4, 8, 9, 14 and 20 molecules \cite{water}. These clusters have similar
electron counts to those of the Li clusters discussed above. Fig. \ref{fig_2}
shows decorrelation plots as a function of block size. All clusters display
the onset of the decorrelation plateau. A comparison of the insets in Figs.
\ref{fig__1} and \ref{fig_2} shows that the value of $\sqrt{\tau}%
=\sigma/\sigma_{0}$ varies significantly less rapidly with system size in
H$_{2}$O clusters. It is striking that all curves in Fig. \ref{fig_2} collapse
onto an essentially universal curve up to five block transformations. Beyond
this block size, however, it appears that other secondary longer-distance
correlations play a noticeable role so that the curves begin to rise again and
settle into new, weakly size-dependent plateaux. If it were not for this
secondary feature, all H$_{2}$O clusters would achieve their plateaux in the
same small number of blocking steps, and the stickiness induced by correlated
sampling would not be important. Because of these secondary features, H$_{2}$O
clusters exhibit quadratic scaling of $\tau$ with $M$, but the influence is
significantly less severe than for Li. \ In fact, if $a+bN_{el}$ describes a
linear fit for the data in the inset for H$_{2}$O, we see that for the cluster
sizes we studied, the constant term in the quadratic expression $\tau=\left(
\sqrt{\tau}\right)  ^{2}=\left(  a+bN_{el}\right)  ^{2}$ dominates. \ This may
explain why the phenomenon was not observed in Ref. \cite{galli}. For Li, the
linear term dominates and adds to the scaling of the total computation time
for the cluster sizes we studied.%

%TCIMACRO{\FRAME{ftbpFU}{3.493in}{2.8781in}{0pt}{\Qcb{(Color online) Same as
%Fig. \ref{fig__1}, except that this figure refers to H$_{2}$O clusters.}%
%}{\Qlb{fig_2}}{h2o_clu.eps}{\special{ language "Scientific Word";
%type "GRAPHIC";  maintain-aspect-ratio TRUE;  display "USEDEF";
%valid_file "F";  width 3.493in;  height 2.8781in;  depth 0pt;
%original-width 6.9444in;  original-height 5.9525in;  cropleft "0";
%croptop "1";  cropright "1";  cropbottom "0";
%filename 'C:/Documents and Settings/Physics_on_Fujitsu/My Documents/_Workspace/MathText/My Published Articles/Decorrelation/h2o_clu.eps';file-properties "XNPEU";}%
%}}%
%BeginExpansion
%EndExpansion

We now provide a possible explanation for the divergence of the
autocorrelation time $\tau$ in ground state computations of Li clusters. We
recall that bulk Li metal contains discontinuities in the momentum density of
the electron gas at the Fermi momentum. Such a sharp cut-off in the momentum
distribution can be expected to generally cause oscillations of electron
configurations involved in the QMC algorithm and thereby slow the statistical
convergence in clusters like those of Li atoms, which must eventually approach
the bulk metallic state with increasing size. These considerations suggest
that smearing or broadening the underlying "Fermi surface" (formally, of
course, there is no Fermi surface in a finite system) in metallic nanoclusters
using a trial wavefunction such as an AGP \cite{agp} or a Pfaffian
\cite{pfaffian} might result in shorter autocorrelation times. Notably, Baroni
and Moroni \cite{baroni} have discussed a theoretical connection between the
autocorrelation time and the spectrum of an auxiliary quantum Hamiltonian,
which might help explain the divergence of the autocorrelation time in
metallic systems.

The integrated autocorrelation time $\tau$ is related to the second largest
eigenvalue $\lambda$ of the transition matrix by $\tau=1+2\lambda/(1-\lambda)$
\cite{liu}. Very recent work regarding the scaling of $\lambda$ in the
Metropolis alorithm demonstrates that for log-concave distributions $|\Phi
_{0}(x)|^{2}$, the scaling of $\tau$ with $M$ is at most $M^{2}$ \cite{novak}.
For Langevin dynamics the divergence of $\tau$ is usually less severe than in
the Metropolis algorithm \cite{roberts}. In principle, the appearance of nodes
in the wavefunction can violate the log-concavity of $|\Phi_{0}(x)|^{2}$ and
might modify the scaling properties of $\tau$. Although our wavefunctions
contain nodes, the fact that we obtain a clear $M^{2}$ scaling of $\tau$
suggests that the $M^{2}$ scaling remains maximal for typical QMC
distributions $|\Phi_{0}(x)|^{2}$ more generally \cite{footnotenodal}.

In conclusion, we have investigated the scaling of the integrated
autocorrelation time $\tau$ in QMC computations of Li and H$_{2}$O clusters.
The quantity $\tau$, which directly yields a measure of sampling correlations
in the calculation, is found to diverge quadratically with system size $M$ in
both cases, more severely for Li than for H$_{2}$O, and becomes increasingly
difficult to calculate for large clusters. We have shown that brute force
decorrelation of samples is simply not feasible for these large clusters. Our
study highlights the importance of correlated sampling in QMC computations of
large clusters and provides a route for obtaining accurate error estimates by
exploiting the scaling properties of $\tau$ with system size, without
requiring explicit decorrelation of samples. The present results should be
relevant for the growing scientific community \cite{liu} interested in the
convergence properties of Markov chains and QMC calculations of properties of
nanoclusters. Further work towards understanding the effects of excitation
gaps on the convergence of the QMC algorithm might be valuable.

We acknowledge useful discussions with Lubos Mitas, Chip Kent, Dario
Bressanini, and Erich Novak, and we thank R. Rousseau and D. Marx for sending
us atomic coordinates for Li clusters. This work was made possible by the
support of the U.S.D.O.E. contracts DE-FG02-07ER46352 and DE-AC03-76SF00098.
We benefited from the allocation of supercomputer time at NERSC and
Northeastern University's Advanced Scientific Computation Center (ASCC).


\begin{thebibliography}{99}
\bibitem {footnoteCorrelations}The underlying algorithm of such simulations
involves Markov chains \cite{liu}. The data generated from a Markov chain are
serially correlated, meaning that the covariance between data elements is
nonzero (i.e., nearby points in the chain are also nearby in configuration
space, not sampled uniformly from the desired distribution).
\bibitem {footnote3}We use only one walker, because this is required for a
study of the decorrelation plateau, defined later.
\bibitem {footnoteSingleGlobalExplanation}We ran preliminary calculations of
small Li clusters using single-electron moves. These calculations revealed
similar modified autocorrelation times \cite{umrigar93b} as global moves.
Also, the QMC step size scales linearly with the inverse of the number of
electrons for the Li clusters we studied, so that the total distance the
electrons move remains the same, whether global or single-electron moves are used.
\bibitem {footnoteEquilibration}We performed a detailed study of Li$_{64}$ and
$\left(  \text{H}_{2}\text{O}\right)  _{20}$ equilibration by running a
massively parallel calculation (1024 processors), assuming the points were
equilibrated (even when they were not), and noticing that the variational
theorem demands that the energy be higher for unequilibrated systems.
Therefore, only when the error bars for our massively parallel run overlapped
the (known) HF energy did we consider our samples to be equilibrated.
\bibitem {lithium}The geometry of Li clusters can be found in \cite{marx}. For
Li$_{64}$ and Li$_{26}$, we have extracted the geometry from bulk BCC Li.
\bibitem {footnoteHF}The use of the HF wavefunction is adequate for our study,
aimed at understanding convergence properties of QMC calculations, rather than
obtaining an accurate value of the correlation energy. Also, by duplicating a
known HF result, we can be certain the system is ergodic.
\end{thebibliography}
\end{document}